\documentstyle[epsfig,aps,pra]{revtex}
\begin{document}

\twocolumn[\hsize\textwidth\columnwidth\hsize\csname
@twocolumnfalse\endcsname

\title{Error-free optical quantum communication}
\author{Dik Bouwmeester}
\address{Centre for Quantum Computation, Clarendon Laboratory,
University of Oxford, \\ Parks Road, OX1 3PU Oxford, United
Kingdom} \maketitle

\begin{abstract}
An optical scheme for the reliable transfer of quantum information
through a noisy quantum channel is proposed. The scheme is
inspired by quantum error-correction protocols, but it avoids the
currently infeasible requirement for a controlled-NOT operation
between single photons. The quantum communication scheme presented
here rejects single bit-flip errors instead of correcting them and
combines quantum-measurement properties of three-particle
entangled (GHZ) states with properties of quantum teleportation.
\end{abstract}


]

The possibility to detect and correct errors in the evolution of a
quantum system has been a most remarkable theoretical discovery
\cite{STE95a,STE95b,SHO95,CAL96,SPR00}. This discovery, and the
subsequent theoretical development of related ideas such as
entanglement purification \cite{BEN96,DEU96} the quantum repeater
\cite{BRI98,DUR99} and fault-tolerant quantum computation
\cite{SHO96,PRE98,STE99}, turned the initial scepticism about
implementing quantum computation and long distance quantum
communication into (modest) optimism. At present, however, no
practical realization of any of these ideas has been achieved in
laboratories. Of particular interest would be an implementation in
quantum optics since this would enable secure quantum cryptography
and quantum communication through optical channels such as optical
fibers. The reason why no such implementation has been realized to
date is that all theoretical schemes are based on controlled-NOT
operations between single particles. For photons this operation
would require a strong non-linear interaction between individual
photons which is extremely difficult to achieve. In this Letter a
scheme is proposed that rejects erroneous transmission of photon
states without using controlled-NOT operations.

\begin{figure}
\centerline{\epsfig{file=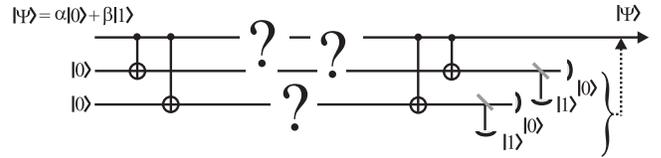,width=8.5cm}} \vspace{0.2cm}
\caption{Traditional scheme for the detection and correction of a
bit-flip error. Using two controlled-NOT operations, an initial
quantum state $| \Psi \rangle$ is entangled with two auxiliary
particles, each initially prepared in the state $|0 \rangle$.
After transmission of the three-particle entangled state through
an area in which an error might occur, indicated by the question
marks, each of the two auxiliary particles becomes the target
particle of a controlled-NOT operation with the initial particle
as the control particle. A final projection measurement on each of
the two auxiliary particles onto the $|0\rangle$, $|1\rangle$
basis uniquely identifies a possible (single) error which can then
be corrected.} \label{Qerror}
\end{figure}

In order to explain the optical scheme we first point out the main
ideas underlying classical and quantum error detection. A
particularly simple classical error detection scheme uses the
transmission of several copies of the bits to be transferred and
requires that the probability of a bit-flip error during
transmission is much smaller than unity. By comparing the copies
of each initial bit after transmission one can determine the
initial bits with high probability. Despite the fact that it is
impossible to copy the state of an unknown quantum state, it is
still possible to use a strategy similar to the classical one.
Consider the state of a two-level system, a qubit, characterized
by
\[ |\Psi \rangle = \alpha|0 \rangle + \beta |1 \rangle \,.\]
In order to make comparison measurements after the state
transmission, and thereby detect errors, we have to encode the
initial qubit onto several particles. If we restrict our attention
to the case that there is a small probability that a bit-flip
error occurs, it is sufficient to encode the initial state onto
the following three-particle entangled state:
\begin{equation}
| \Psi \rangle_{123}= \frac{1}{\sqrt{2}} \left( \alpha
|000\rangle_{123} + \beta |111\rangle_{123} \right) \,.
\end{equation}
The left-hand side of Fig.~\ref{Qerror} indicates how this
encoding is obtained using two controlled-NOT operations with the
initial qubit as control qubit and two auxiliary particles
initially prepared in state $|0\rangle$ as target qubits.

After transmission of the three-particle entangled state through a
``noisy'' quantum channel one can retrieve the initial qubit using
the comparison measurements indicated on the right-hand side of
Fig.~\ref{Qerror}. The measurements consist again of
controlled-NOT operations, followed by detection of the two
auxiliary particles in the $|0\rangle$, $|1\rangle$ basis. The
detection acts as a parity check between the two particles on
which the controlled-NOT operation acts: a $|0\rangle$ outcome
indicates that in each term of the entangled state the two
particles are the same, {\it i.e.}, 00 or 11,  a $|1\rangle$
outcome indicates that they are opposite, {\it i.e.}, 01 or 10. If
no error occurred during the transmission, both auxiliary
particles should be detected in the state $|0\rangle$. However, if
a bit-flip error occurred for the initial particle, and not for
the other two, both auxiliary particles will be detected in the
state $|1\rangle$. In the case that an error occurred on one of
the auxiliary particles, and not on the remaining two, the
corresponding particle will be detected in state $|1\rangle$ and
the remaining auxiliary particle in state $|0\rangle$. After
identification, a possible error can be corrected.

Crucial for the error detection/correction scheme is that under
the assumption that at the most one error occurs the parity-check
measurements project the transmitted entangled state onto only
four possible outcomes, namely no error, or one error on one of
the three particles. Therefore, although during the transmission
through the noisy quantum channel any qubit-rotation error can
occur, the final state is quantized to contain either a full
bit-flip error or no error.

\begin{figure}
\centerline{\epsfig{file=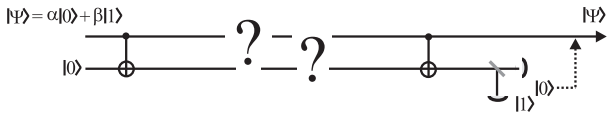,width=8cm}} \vspace{0.2cm}
\caption{Scheme for bit-flip error rejection. One auxiliary
particle is sufficient in order to detect an error, without
revealing on which particle the error occurred.} \label{exclude}
\end{figure}

If more than one error occurred the error-correction scheme is not
useful. Therefore, it is crucial that the probability for an error
on each particle is much smaller than unity
($P_{\mbox{\footnotesize{error}}}\ll 1$). Under this condition it
is reasonable to consider for optical quantum communication
purposes a simplified scheme that rejects transmissions that
contain an error instead of identifying a specific error and
correcting for it. Such a simplified scheme requires only one
auxiliary particle as shown in Fig.~\ref{exclude}. If the parity
check measurement yields the $|0\rangle$ result, no error took
place, or, with the very small probability
$P_{\mbox{\footnotesize{error}}}^2$, a fatal double-error took
place. If the measurement yields the $|1\rangle$ result, a single
error occurred for one of the two particles and the transmission
is invalidated.

\begin{figure}
\centerline{\epsfig{file=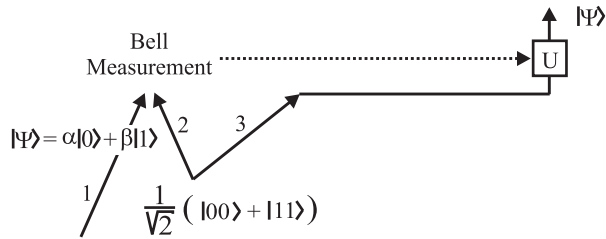,width=8cm}} \vspace{0.2cm}
\caption{Schematic drawing of the quantum teleportation protocol.
The transmission of the unknown quantum state $|\Psi \rangle$ of
particle 1 is broken down into the distribution of an auxiliary
pair of entangled particles (2 and 3), a Bell-state measurement on
particles 1 and 2 ({\it i.e.}, a projection onto a complete basis
of maximally entangled particles), and the transfer of classical
information (the outcome of the Bell-state measurement). After
receiving the classical information, the relation of the state of
particle 3 to the initial state $|\Psi \rangle$ is fully
determined. The initial state can therefore be recovered by a
well-defined unitary transformation $U$ on particle 3.}
\label{tel}
\end{figure}

To present our error-free optical quantum communication scheme we
note that the controlled-NOT operation in the preparation step of
the schemes shown in Figs.~\ref{Qerror} and ~\ref{exclude} is used
in order to encode an {\em{arbitrary}} initial quantum state onto
a multi-particle entangled state.  It is, however, not necessary
to be able to encode an arbitrary input state. According to the
teleportation scheme \cite{BEN93}, illustrated in Fig.~\ref{tel},
the transmission of an arbitrary quantum state can be decomposed
into the transmission of a {\em known} entangled state, a local
Bell-state measurement and the transmission of classical
information. Therefore, in order to establish error-free quantum
communication, it is sufficient to be able to exclude erroneous
transmission of one of the particles of a fixed entangled state.

\begin{figure}
\centerline{\epsfig{file=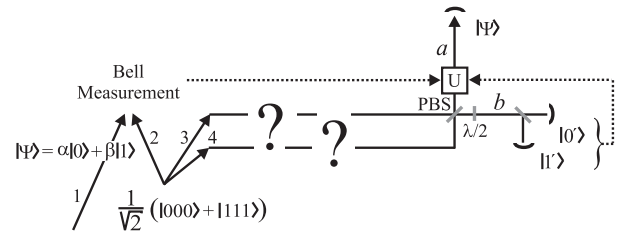,width=8cm}} \vspace{0.2cm}
\caption{Scheme for error-free quantum-state transmission without
controlled-NOT operations. In order to transfer a quantum state it
is sufficient to restrict the use of a quantum channel for the
transmission of one of an entangled pair of particles (see
Fig.~\ref{tel}). In order to reject erroneous transmissions, a
three-particle entangled state is used. Two of the three entangled
particles are sent through the ``noisy'' quantum channel. A
parity-check measurement on particles 3 and 4 identifies an
error-free transmission and is obtained by using a polarizing beam
splitter (PBS) followed by a coincidence detection of one particle
in arm $a$ and the other in arm $b$. The measurement in arm $b$
must be such that the remaining two particles are projected onto a
well-defined two-particle entangled state. This is achieved by
performing the measurement in the linear basis rotated
45$^{\circ}$ with respect to the $|0 \rangle$, $|1 \rangle$ basis.
After the result of the measurement on the particle in arm $b$ is
known, the remaining particles (one to be detected in arm $a$ and
particle 2) are guaranteed to be in a well-defined entangled state
and can be used for error-free quantum teleportation of the state
of particle 1 or for quantum cryptography.} \label{new1}
\end{figure}

Consider a pair of entangled photons in the state
\begin{equation}
| \Psi \rangle_{23}= \frac{1}{\sqrt{2}} \left( |0\rangle_2
|0\rangle_3 +|1\rangle_2 |1\rangle_3 \right) \,.
\end{equation}
To be able to detect errors on the transmission of, say, photon 2,
the preparation scheme shown on the left-hand side of
Fig.~\ref{exclude} would produce the state
\begin{equation}
| \Psi \rangle_{234}= \frac{1}{\sqrt{2}} \left( |0\rangle_2
|00\rangle_{34} +|1\rangle_2 |11\rangle_{34} \right) \,.
\label{initial}
\end{equation}
Since state (\ref{initial}) is a well-defined state, the use of
the controlled-NOT operation is no longer necessary, as shown in
the left-hand side of Fig.~\ref{new1}.

The right-hand side of Fig.~\ref{new1} illustrates how the
controlled-NOT operation for  parity checking can also be avoided
by using a polarizing beam splitter and a coincidence detection
measurement in an appropriate basis. If a bit-flip error occurred
for one of the two transmitted photons, both photons will exit the
polarizing beam splitter in the same output arm. Therefore no
coincidence will be observed between the detectors in arm $a$ and
$b$, and the transmission will be invalidated.

If no error occurred the state after the polarizing beam splitter
will have one photon in each output arm, indicating that the two
outgoing photons have the same polarization relation as when
initially prepared, {\it i.e.}, the polarizations are parallel in
each term of the entangled state. The detection scheme proceeds by
detecting the particle in arm $b$ in the basis
\begin{equation}
|0^{\prime} \rangle = \frac{1}{\sqrt{2}} ( |0\rangle + |1\rangle )
\,,\,\,\,\, |1^{\prime} \rangle = \frac{1}{\sqrt{2}} ( |0\rangle -
|1\rangle) \,.
\end{equation}
The specific measurement outcome corresponds to a projection of
the remaining particles, provided a particle is present in arm
$a$, onto one out of two well-defined pure two-particle entangled
states:
\begin{eqnarray}
|0^{\prime} \rangle_{b} & \rightarrow & \frac{1}{\sqrt{2}} \left(
|0\rangle_2 |0\rangle_{a} +|1\rangle_2 |1\rangle_{a} \right) \\
|1^{\prime} \rangle_{b} & \rightarrow & \frac{1}{\sqrt{2}} \left(
|0\rangle_2 |0\rangle_{a} - |1\rangle_2 |1\rangle_{a} \right)\,.
\end{eqnarray}
The teleportation procedure can now be completed by a Bell-state
measurement and the transfer of classical information as
illustrated in Fig.~\ref{new1}. The unitary transformation $U$ is
determined by the measurement result in arm $b$ and by the result
of the Bell-state measurement.

One might be alarmed by the fact that the photon in arm $a$ has
still to be detected in order to complete the error-free
transmission scheme. This will in practical applications imply the
destruction of the photon, although absorption-free detection of
single-photons has experimentally been demonstrated \cite{NOG99}.
The anxiety to lose the photon before being able to use it is
unjustified, at least for applications in quantum cryptography and
other quantum communication protocols, since the detection of the
photon is an integral part of all such applications. In fact, any
realistic single-photon communication scheme needs a final
verification step to guarantee that the fragile photon survived
the transmission. The detection of the photon, therefore, plays
the double role of enabling a projection onto a pure entangled
state for photon 2 and the photon in arm $a$ , as well as
exploring this entanglement for quantum cryptography or for
quantum communication purposes.

We will now briefly concider the importance of the present scheme
for the security of quantum cryptography \cite{SPR00,EKE91,EKE92}.
A major threat to the security of quantum cryptography based on
entangled state distribution is that the entanglement distribution
established between the two users is in general not perfect. The
users have to agree on a minimum level of security corresponding
to a certain purity of the distributed entanglement. There are
many technical reasons for the accidental loss of entanglement
(transmission through a noisy quantum channel) but one should also
take into account the possibility that a third party is
deliberately introducing small errors on what could otherwise be a
perfect quantum channel (transmission through a {\em hostile}
quantum channel). Since the users cannot distinguish between a
noisy and a hostile quantum channel they are forced to use
imperfect entangled photon pairs for the distribution of a secret
key at the risk of leaking some of the secret key to a third
party. The scheme presented in this Letter solves both the
``noisy'' and the ``hostile'' quantum-channel problem (up to
second-order errors). The users will obtain entangled photon pairs
with the same purity as the initially created three-particle
entanglement despite the fact that the quantum channel is
imperfect. Any accidental or deliberately induced errors on the
transmission will be systematically rejected. Note that a quantum
key-distribution protocol based on the proposed scheme for
error-free entanglement distribution will involve one more
classical communication step in addition to the familiar classical
communication steps, namely the transfer of the detection result
of the photon in arm $b$.

\begin{figure}
\centerline{\epsfig{file=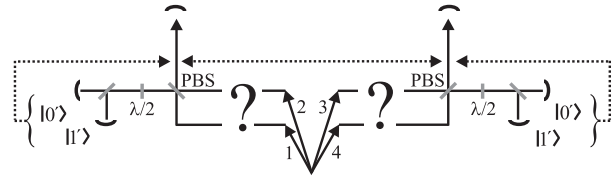,width=8cm}} \vspace{0.2cm}
\caption{Generalization of the reliable transmission of one of an
entangled pair of particles to the reliable distribution of both
entangled particles. Starting with a four-particle entangled
state, an error-free (provided that not more than one error
occurred in each of the two distribution arms) entangled pair of
particles can be obtained.} \label{new2}
\end{figure}

We have restricted our attention to the reliable transmission of
half of an entangled pair. The scheme can easily be extended to
error-free distribution of both particles of an entangled pair as
illustrated in Fig.~\ref{new2}.
So far, we only discussed bit-flip errors. In addition there could
be phase errors. Error detection/correction schemes have been
developed for correcting general errors consisting of both
bit-flip and phase errors \cite{CAL98}. Such schemes involve at
least three/four auxiliary particles and more elaborate
preparation and detection procedures. It remains to be seen
whether an all-optical scheme is possible to reject both bit-flip
and phase errors.

Currently we are working towards an experimental realization of
the scheme presented in this Letter. At first glance, it seems
that previous experiments on quantum teleportation \cite{BOU97}
and three-photon entanglement \cite{BOU99} provide all the
necessary techniques to implement the scheme. Unfortunately, the
source for three-particle entanglement reported in
Refs.~\cite{BOU99,PAN00} is based on a post-selection detection
method which filters out the appropriate three-photon entanglement
from a variety of other photon states. Since in the proposed
scheme two of the three entangled photons are re-combined on a
beam splitter before detection, the post-selection cannot be
applied. What seems to be needed is a three-photon source which
produces three spatially separated outgoing photons in a genuine
three-photon Greenberger-Horne-Zeilinger state. Methods to encode
information on more than one degrees of freedom of single photons
(polarization and momentum) might also be employed to achieve an
experimental realization \cite{POP,BOS98,ZUK91}.

Finally we point out the generality of the two main ideas of this
Letter. First, errors on the evolution of entangled states can be
detected by starting with higher-order entangled states. Second,
it appears that an experimental implementation of the
controlled-NOT operation, or any other universal quantum gate, is
not crucial for an experimental demonstration of the essence of a
variety of quantum-communication protocols.

\section*{acknowledgement}
I am very grateful to John Rarity, Sandu Popescu, Andrew Steane,
Antia Lamas Linares, Christian Mikkelsen and Michael Seevinck for
stimulating discussions and useful suggestions. This work was
supported by the EPSRC GR/M88976 and the European QuComm
(ISI-1999-10033) projects.


\begin{thebibliography}{99}
\bibitem{STE95a} A.~Steane, Phys. Rev. Lett. {\bf 77}, 793
(1995).
\bibitem{STE95b} A.~Steane, Proc. R. Soc. Lond. A, {\bf 452}, 2551 (1995).
\bibitem{SHO95} P.W.~Shor, Phys.
Rev. A, {\bf 52}, R2493 (1995).
\bibitem{CAL96} A.R.~Calderbank and P.W.~Shor, Phys. Rev. A, {\bf 54},
1098 (1996).
\bibitem{SPR00} For an overview see {\em The Physics of Quantum Information},
Eds. D.~Bouwmeester, A.~Ekert, A.~Zeilinger, Springer-Verlag,
Berlin Heidelberg New York (2000).
\bibitem{BEN96} C.H.~Bennett, G.~Brassard, S.~Popescu, B.~Schumacher, J.A.~Smolin, W.K.~Wootters,
Phys. Rev. Lett. {\bf 76}, 722-725 (1996).
\bibitem{DEU96} D.~Deutsch, A.~Ekert, R.~Jozsa, C.~Macchiavello, S.~Popescu,
A.~Sanpera, Phys. Rev. Lett. {\bf 77}, 2818 (1996).
\bibitem{BRI98}
H.-J.~Briegel, W.~D\"ur, J.I.~Cirac, and P.~Zoller, Phys. Rev.
Lett. {\bf 81}, 5932 (1998).
\bibitem{DUR99}
W.~D\"{u}r, H.-J.~Briegel, J.I.~Cirac, and P.~Zoller, Phys. Rev. A
{\bf 59}, 169 (1999); ibid. {\bf 60}, 729 (1999).
\bibitem{SHO96}  P.W.~Shor, {\em Proc. 37th Symp. on Foundations of Computer Science}, (Los
Alamitos, CA: IEEE Computer Society Press), pp15-65 (1996).
\bibitem{PRE98} J.~Preskill, Proc. Roy. Soc. Lond. A, {\bf 454}, 469 (1998).
\bibitem{STE99} A.M.~Steane, Nature, vol. {\bf 399}, 124-126 (1999).
\bibitem{BEN93} C.H.~Bennett, G.~Brassard, C.~Cr\'{e}peau,
R.~Jozsa, A.~Peres, W.K.~Wootters,  Phys. Rev. Lett. {\bf 70},
1895 (1993).
\bibitem{NOG99} G.~Nogues, A.~Rauschenbeutel, S.~Osnaghi, M.~Brune, J.M.~Raimond and S.~Haroche,
Nature {\bf 400}, 239 (1999).
\bibitem{EKE91} A.K.~Ekert, Phys. Rev. Lett. {\bf 67}, 661 (1991).
\bibitem{EKE92} A.K.~Ekert, J.G.~Rarity, P.R.~Tapster, G.M.~Palma,
Phys.~Rev.~Lett. {\bf 69} 1293 (1992).
\bibitem{CAL98} A.R.~Calderbank, E.M.~Rains, P.W.~Shor,
J.A.~Sloane, IEEE Trans. on Information theory, {\bf 44}, 1369
(1998).
\bibitem{BOU97} D.~Bouwmeester, J.-W.~Pan, K.~Mattle,
M.~Eible, H.~Weinfurter, A.~Zeilinger,  Nature {\bf 390}, 575-579
(1997).
\bibitem{BOU99} D.~Bouwmeester, J.-W.~Pan, M.~Daniell, H.~Weinfurter, A.~Zeilinger,
Phys. Rev. Lett. {\bf 82}, 1345 (1999).
\bibitem{PAN00} J.-W.~Pan, D.~Bouwmeester, M.~Daniell, H.~Weinfurter,
A.~Zeilinger, Nature {\bf 403}, 515 (2000).
\bibitem{POP} S.~Popescu, LANL E-print quant-ph 9501020.
\bibitem{BOS98} D.~Boschi, S.~Branca, F.~De~Martini, L.~Hardy, S.~Popescu,
Phys. Rev. Lett. {\bf 80}, 1121 (1998).
\bibitem{ZUK91} M.~Zukowski, Phys. Lett. A 157, 198 (1991).

\end{thebibliography}
\end{document}